\documentclass[12pt,preprint]{aastex}
\usepackage{amssymb}

\begin{document}

\title{Resolved Mid-IR\ Emission in the Narrow Line Region of NGC 4151}
\author{James T. Radomski, Robert K. Pi\~{n}a, Christopher Packham, Charles M.
Telesco}
\affil{Department of Astronomy, University of Florida, Gainesville, FL
32611, USA}

\and
\author{James M. De Buizer}
\affil{Cerro Tololo Inter-American Observatory (CTIO), National Optical Astronomy Observatory, Casilla 603, La Serena, Chile}

\and 
\author{R. Scott Fisher}
\affil{Gemini Observatory, Northern Operations Center, 670 N. A'ohoku Place, Hilo, HI  96720, USA}

\and
\author{A. Robinson}
\affil{Department of Physical Sciences, University of Hertfordshire, College Lane, Hartfield, Hertfordshire, AL10 9AB, UK }

\begin{abstract}
We present subarcsecond resolution mid infrared images of NGC 4151 at 10.8$%
\micron$ and 18.2$\micron$. These images were taken with the University of
Florida mid-IR camera/spectrometer OSCIR at the Gemini North 8 m telescope.
We resolve emission at both 10.8$\micron$ and 18.2$\micron$ extending $%
\thicksim 3.5\arcsec$ across at a P.A. of $\thicksim 60\degr$. This
coincides with the the narrow line region\ of NGC 4151 as observed in [OIII]
by the Hubble Space Telescope. The most likely explanation for this extended
mid-IR emission is dust in the narrow line region heated by a central
engine. We find no extended emission associated with the proposed torus and
place an upper limit on its mid-IR size of $\lesssim 35$ pc. 
\end{abstract}


\section{Introduction}

NGC 4151 is one of the nearest (13.2 Mpc, H$_{o}=75$ km s$^{-1}$ Mpc$^{-1}$)
and best studied active galactic nuclei (AGN). The nucleus hosts a highly
variable continuum and line emission source. Continuum variability, first
reported by Fitch et al. (1967), has been observed at several wavelengths
including X-ray (Papadakis et al. 1995), UV (Clavel et al. 1990), and
optical (Lyutyi 1972). Classified as a Seyfert 1.5 by Osterbrock \& Koski
(1976), NGC\ 4151 displayed characteristics of a Seyfert 2 (Penston \& P{$%
\acute{e}$}rez 1984) during a low luminosity state in 1984, and at a later
date characteristics of a Seyfert 1 (Ayani \& Maehara 1991).

The mid-infrared emission from NGC\ 4151 has been suggested to arise from
either thermal emission from dust grains or synchrotron emission.
Discussions of the thermal vs. nonthermal origin of the infrared emission in
NGC 4151 can be found in Rieke \& Lebofsky (1981), Edelson \& Malkan (1986),
Carelton et al. (1987), Edelson et al. (1987), and de Kool \& Begelman
(1989). A direct method to investigate the origin of the mid-IR emission
mechanism, as proposed by Neugebauer et al. (1990) (hereafter N90), is a
measurement of the size of the emitting region. They suggest that a
nonthermal self-absorbed synchrotron source would be $<1$mas, and hence
unresolvable. However, if the mid-IR emission is due to heated dust grains,
the size of the region would be $>0.1\arcsec$.

Observations show that the mid-IR emission in NGC 4151 is compact.
Comparison between 60$\arcsec$ resolution IRAS 12 $\micron$ flux density
measurements and 6$\arcsec$ resolution ground-based 10.6 $\micron$
measurements agree to within $\thicksim 6\%$ (Edelson et al. 1987). Mid-IR
observations by Rieke \& Low 1972; Rieke \& Lebofsky 1981; Ward et al. 1987
also did not detect any extended emission with resolutions $\gtrsim 6\arcsec$%
. Observations by ISO (Infrared Space Observatory; Rodriquez-Espinosa et al.
1996 - hereafter RE96) show a strong warm dust component in NGC\ 4151 and
suggest a thermal origin from a geometrically and optically thick dusty
torus and/or a dusty narrow line region (NLR). These observations however
were at a resolution of $180\arcsec$. Using a technique of near-simultaneous
north-south scans at 2.2 $\micron$ and 11.2 $\micron$, N90 was able to
resolve the 11.2 $\micron$ emitting region to be 0\farcs16$\pm 0$\farcs04.
However this technique measured the size in only one spatial direction and
was insufficient to explore the mid-IR morphology of the circumnuclear
region. In addition, these north-south scans could not investigate the NLR
of NGC 4151 which is primarily orientated in an east-west direction. In this
paper we present high resolution mid-IR imaging which, to the best of our
knowledge, resolves the inner NLR\ of NGC 4151 for the first time at 10 $%
\micron$ and 18 $\micron$.

\section{Observations and Data Reduction}

Observations of NGC 4151 were made on 7 May 2001 using the University of
Florida mid-infrared camera/spectrometer OSCIR on the Gemini North 8-m
telescope. OSCIR uses a Rockwell 128 x 128 Si:As Blocked Impurity Band (BIB)
detector. On Gemini North, OSCIR has a plate scale of 0\farcs089 pixel$^{-1}$%
, corresponding to a field of view of \ 11\farcs4 x 11\farcs4. Images were
obtained in the N ($\lambda _{o}$=10.8 $\micron$,\ $\Delta \lambda $=5.2 $%
\micron$) and IHW18 ($\lambda _{o}$=18.2 $\micron$,\ $\Delta \lambda $=1.7 $%
\micron$) filters using a standard chop/nod technique to remove sky
background and thermal emission from the telescope. The chopper throw was 15$%
\arcsec$ in declination at a frequency of 3 Hz and the telescope was nodded
every 30 seconds.

NGC 4151 was observed for a total chopped integration time of 360 seconds at
10.8 $\micron$ and 480 seconds at 18.2 $\micron$. Observations of $\beta $\
Gem were taken for flux calibration and as a measure of the telescope point
spread function (PSF). Measurements of other calibration stars throughout
the night showed flux calibration variations of less than 5\% at 10.8 $%
\micron$ and less than 10\% at 18.2 $\micron$. Absolute calibration of\ $%
\beta $\ Gem was achieved using a spectral irradiance model by Cohen (1995)
adjusted for filter and atmospheric transmission. The calibration value and
FWHM were also color corrected to account for the different spectral slope
of $\beta $\ Gem versus NGC 4151 as observed within our N and IHW18 filters.
The measured color corrected FWHM of $\beta $\ Gem was 0\farcs53 at 10.8 $%
\micron$and 0\farcs58 at 18.2 $\micron$ based on a 60 second chopped
integration. Short integrations of $\beta $\ Gem were sufficient for
comparison to NGC 4151 due to the stability of the OSCIR/Gemini PSF.
Observations of several standard stars including $\beta $\ Gem, $\mu $ Uma,
and $\gamma $ Aql showed variations in the FWHM of $<$ 6\% throughout the
night. Finally, observations of NGC 4151 showed no change in structure when
divided into increments of time equal to that of the PSF (60s).

OSCIR\ was mounted on the telescope with the Gemini instrument rotator
oriented such that north was up and east was left on the detector array. In
post-processing, images of the PSF star $\beta $\ Gem were ``unrotated''
-17.4$\degr$ and -22.9$\degr$ \ at 10.8 $\micron$ and 18.2 $\micron$
respectively to match the position angle of the Gemini North telescope 
\textit{pupil }as projected on the detector array when NGC 4151 was
observed. This is necessary to correctly account for the rotation of the
telescope pupil with respect to OSCIR (during an observation or when
changing pointing) due to the alt-az mount of the Gemini North telescope. In
addition, PSF images were rotationally ``smeared'' to account for the slight
rotation of the pupil ($\lesssim $ 4$\degr$) during the exposure times of
NGC 4151.

Flux density maps were created by convolving images at 10.8 $\micron$ with
the 18.2 $\micron$ PSF and vice-versa to attain the same resolution at both
wavelengths. Color temperature and emission optical depth maps were
calculated based on the ratio of these images. Since no astrometric
calibration was performed due to the limited field of view of OSCIR, the
peak flux of the convolved 10.8 $\micron$ image was aligned to coincide with
peak flux of the convolved 18.2 $\micron$ image. Temperature and emission
optical depth were then calculated using the optically thin approximation $%
F_{\nu }=\Omega \tau B_{\nu }(T)$, where $F_{\nu }$ is the observed flux
density at frequency $\nu $, $\Omega $ is the solid angle of each pixel, $%
\tau $ is the emission optical depth, and $B_{\nu }(T)$ is the Planck
function evaluated at frequency $\nu $\ and temperature T. The structure of
the temperature map was highly dependent on the alignment of the two
convolved images. In order to determine the errors due to alignment, a Monte
Carlo simulation was done by shifting the two convolved images with respect
to each other up to 0.1$\arcsec$ in all directions. Temperature values were
most stable in the core varying $\pm $5 K, with variations of $\pm $15 K
further out. The frequency dependence of dust grain emission efficiency in
the mid-IR was approximated as $Q(v)\propto v^{1}$. A steeper power law
dependence such as $Q(v)\propto v^{2}$ would decrease the calculated
temperatures $\thicksim 15\%$ and correspondingly increases the emission
optical depth by a factor of $\thicksim 3.4$.

\section{Results}

NGC 4151 shows extended emission $\thicksim 3\farcs5$ across oriented in an
approximately east-west direction. Perpendicular to the extended emission
(roughly north-south) the galaxy remains unresolved based our limiting
resolution of $\thicksim 0\farcs53$ - $0\farcs58$ at 10.8 $\micron$ and 18.2 
$\micron$ respectively. Figure 1 shows our N-band image of the central $%
\thicksim 400$ pc ($\thicksim 6\arcsec$) of NGC 4151. Subtraction of the
unresolved (PSF) component scaled to $100\%$ of the peak results in a hole
at the center of the residual emission. Since this may represent an
over-subtraction of the unresolved component, we also show the residuals
after subtraction of the PSF scaled to $90\%$, $80\%$, and $70\%$ of the
peak of NGC 4151. Figure 2 shows a similar comparison at 18.2 $\micron$.
Both figures clearly show extended mid-IR emission on a much larger scale
than previously measured.

The extent and position angle ($\sim 60\degr$) of the extended emission is
coincident with the NLR as observed by Evans et al. (1993) and Kaiser et al.
(2000) using HST. This NLR was resolved at [OIII] $\lambda $5007\ into a
number of emission line clouds distributed in a biconical structure oriented
along a PA = 60$\degr\pm 5$, with an opening angle of 75$\degr\pm 10$ (Evans
et al. 1993). NGC\ 4151 also contains a radio jet (Johnston et al. 1982;
Pedlar et al. 1993; and Mundell et al. 1995) which extends along a slightly
different PA\ ($\sim 77\degr$)\ than the [OIII] emission. Figure 3 shows our
mid-IR\ images overlaid on the [OIII] and radio maps in the central $\sim 6%
\arcsec$. The extended mid-IR\ emission strongly coincides with the [OIII]
emission out to a distance of $\sim $100 pc from the nucleus on either side.

As previously mentioned, north-south scans by N90 measure the 11.2 $\micron$
emitting region to be 0\farcs16$\pm 0$\farcs04 or $\thicksim 10$ pc. We see
no extended emission in this direction and can only place an upper limit on
the mid-IR size of $\lesssim 35$pc based on our resolution limit of $%
\thicksim 0\farcs53$ - $0\farcs58$. Scaling the PSF to $100\%$ of the peak
of NGC 4151 results in an upper limit of the unresolved component of $\leq
73\%$ of the total emission at 10.8 $\micron$ and 18.2 $\micron$ and a lower
limit to the extended component of $\geq 27\%$. Table 1 shows our mid-IR
flux density measurements.

\section{Analysis and Discussion}

Based on the conclusion from N90 that a nonthermal self-absorbed synchrotron
source would be $<1$mas, and hence unresolvable, our results are consistent
with a thermal origin of the extended mid-IR emission. The re-radiation by
dust grains heated by either stars or an AGN may result in this extended
thermal mid-IR\ emission. However, processes such as fine structure emission
lines may also produce extended mid-IR emission. Several mid-IR fine
structure lines were observed by Sturm et al. (1999) using ISO. Four of
these emission lines fall within our broadband N and IHW18 filters and may
contribute to this mid-IR emission. These emission lines are the 8.99 $%
\micron$ [ArIII], 10.51 $\micron$ [SIV], 12.81 $\micron$ [NeII], and 18.71 $%
\micron$ [SIII]. However based on flux measurements from Sturm et al.
(1999), these emission lines contribute $<10\%$ of the extended emission we
observe at 10.8 $\micron$ and 18.2 $\micron$.

Several mechanisms can contribute to thermal dust emission in the mid-IR.
These include shock heating, in-situ star formation, dust in the NLR heated
by the central engine, and a dusty torus. Each is considered below in the
context of the mid-IR emission we detect in NGC\ 4151.

\subsection{Dust Heated in Shocks}

Shock heating of dust grains caused by the radio jet may contribute to
mid-IR emission in NGC 4151. Mid-IR emission from shocks may be produced by
either direct collisions between the plasma and dust particles (Draine 1981)
or absorption of UV\ radiation produced by the postshock cooling plasma
(Dopita \& Sutherland 1996). However, radial velocity measurements of Kaiser
et al. (2000) show no correspondence between velocity or velocity dispersion
and the positions of the radio knots of NGC 4151. Models by Crenshaw et al.
(2000) also do not require strong shocks to explain the kinematics of the
NLR. Shocks may still contribute to the NLR as suggested by Contini et al.
(2002), but are not considered to be a major source of ionization. Recent
[Fe II] observations by Turner et. al. (2002) also support this concept.
Thus shocks heating, though not ruled out, is not considered as a major
source of the mid-IR\ emission we observe.

\subsection{Star Formation}

Mid-IR emission has been observed to arise from star formation in the
central regions of many galaxies (e.g. Telesco 1988). The mid-IR provides an
excellent trace of HII star forming regions whose emission peaks at far-IR
wavelengths (50-200 $\micron$). Observations by Engargiola et al. (1988) at
155 $\micron$ show extended emission ($>48\arcsec$) primarily in an
east-west direction. ISO far-IR\ observations by RE96 measure this emission
as a cold dust component (36K) consistent with dust heated in HII regions
(Telesco et al. 1980). However, observations by P{$\acute{e}$}rez-Fournon \&
Wilson (1990) in H$\alpha $ show these HII regions exist in an elliptical
galactic bar at a radius of $\thicksim 50\arcsec$ ($\thicksim 3$ kpc) from
the nucleus. Thus these star forming regions cannot contribute to the mid-IR
emission we observe within our $\thicksim 11\arcsec$\ field of view.

Star formation in the circumnuclear region of NGC 4151 has been
characterized using the strength of polycyclic aromatic hydrocarbon (PAH)\
emission. Galaxies with strong nuclear star formation also feature strong
PAH emission. This PAH\ emission however is found to be weak or absent in
AGNs with weak star formation (Roche et al. 1991; Genzel et al. 1998). In
the case of NGC\ 4151, Roche \& Aitken (1985) and Imanishi et al. (1998)
failed to detect PAH emission at 11.4 $\micron$ and 3.3 $\micron$
respectively with their $\thicksim 4\arcsec$ apertures. Further observation
by Sturm et al. (1999) also failed to detect any PAH emission at 11.2 $%
\micron$, 8.7 $\micron$, 7.7 $\micron$, or 6.2 $\micron$. Thus the mid-IR
emission we observe on a scale of $\thicksim 3.5\arcsec$ is unlikely to be
associated with significant star formation.

\subsection{Dusty Narrow Line Region}

The most likely explanation for the ``extended'' mid-IR morphology in NGC
4151 is emission from a dusty NLR (Rieke et al. 1981; RE96). Dust in this
region has a direct view of the central engine and hence can be heated
resulting in extended mid-IR emission. The emission we observe follows the
NLR as delineated by the [OIII] observations of Kaiser et al. (2000),
lending support to this concept. Mid-IR emission coincident with [OIII] NLR
emission has also been observed in other galaxies such as NGC 1068 (Braatz
et al. 1993, Cameron et al. 1993) and Cyg A (Radomski et al. 2001, 2002). In
both galaxies dust heated by the central engine most likely contributes to
this emission.

In order to explore the possibility of central heating we calculated color
temperatures based on the ratio of our 10.8 $\micron$ and 18.2 $\micron$
images. Temperature and emission optical depth maps from simple radiative
transfer analysis provide a good first-order estimate of the sources of
grain heating as well as the relative density of warm grains
(Tresch-Fienberg et al. 1987). Figure 4 shows our temperature and emission
optical depth maps. We calculate color temperatures ranging from $\thicksim
185\pm 5$ K in the core to $\thicksim 165\pm 15$ K within the NLR ($%
r\thicksim 100$ pc), consistent with the warm dust component (170 K) as
measured by ISO (RE96). The emission optical depth shows the density of
these dust grains is enhanced along the direction of the NLR. Assuming a
simple uniform dust distribution, a first-order determination of the size of
the region that could be heated by a central source can be made. Given that
dust grains primarily absorb UV-optical radiation and re-emit in the
infrared, the equilibrium temperature of dust in a strong UV field is given
by (Sellgren et al. 1983)

\begin{equation}
T\sim \left( \frac{L_{UV}}{16\pi R^{2}\sigma }\frac{Q_{UV}}{Q_{IR}}\right)
^{1/4}
\end{equation}
In the above equation, T is the dust temperature, $L_{UV}$ is the UV
luminosity of the central source, $R$ is the radius from the source in
parsecs, $\sigma $ is the Stefan-Boltzman constant, and $Q_{UV}/Q_{IR}$ is
the ratio of the Planck averaged UV absorption coefficient to the infrared
emission coefficient. Values of $Q_{UV}/Q_{IR}$ are dependent on the dust
grain size and composition and are given by Draine \& Lee (1984), Laor \&
Draine (1993), and Weingartner \& Draine (2001) for graphite and ``smoothed
astronomical'' (SA) silicate.

The observed UV-optical luminosity of NGC 4151 is $\thicksim 10^{10}\ $L$_{%
\sun}$ (Penston et al. 1990). Given this luminosity, in order to heat dust
to the observed temperature of $\thicksim 165\pm 15$ K at a distance of $%
\thicksim 100$ pc, the inner NLR\ should be composed of $0.004$ $\micron$
graphite grains or $0.001$ $\micron$ SA silicate grains. These grain sizes
fall near or below the limit of classical interstellar dust grains which
range from 0.003 $\micron$ - 1 $\micron$ (Draine \& Lee 1984). They also are
much smaller than the estimated grains sizes used for the centrally heated
NLR models of NGC 1068 ($\sim 0.05$ $\micron$; Cameron et. al. 1993) and Cyg
A ($\sim 0.1$ $\micron$; Radomski et. al. 2002). However Penston et al.
(1990) proposed that the continuum emission in NGC 4151 is inherently
anisotropic and that the ionizing luminosity as seen within the extended NLR
(1-2 kpc) may be on order of 13 times greater than that observed from
Earth. Subsequent models by Schulz \& Komossa (1993), Yoshida \& Ohtani
(1993), and Robinson et al. (1994) also derive values for the anisotropy as
high as 3-10. Thus a better estimate of the luminosity in the NLR may be $%
\sim 10^{11}\ $L$_{\sun}$. This higher luminosity would increase the
calculated grain sizes by an order of magnitude ($0.04$ $\micron$ graphite
or $0.01$ $\micron$ SA silicate), resulting in sizes more consistent with
classical interstellar dust and grain size estimates in the NLRs of NGC 1068
and Cyg A. Thus to first order, assuming that the luminosity in NGC\ 4151 is
anisotropic, the extended mid-IR emission is consistent with heating of dust
in the NLR from a central engine.

\subsection{Dusty Torus}

Another source of mid-IR emission in NGC 4151 may be from a dusty torus
(RE96). Emission from a dusty torus may dominate the unresolved mid-IR
component in NGC\ 4151. It is considered that we view this disk/torus
through a line-of-sight passing near the boundary edge (Cassidy \& Raine
1996). Assuming the torus lies perpendicular to the NLR, it's major axis
would be oriented in an approximately north-south direction. We see no
extended emission in this direction and can only place an upper limit on the
mid-IR size of the torus of $\lesssim 35$ pc based on our resolution limit
of $\thicksim 0\farcs53$ - $0\farcs58$. This is consistent with the
polarimetry observations and subsequent modelling of Ruiz et al. (2002)
which suggest that the torus size in NGC 4151 is $\thickapprox $ 30 pc. A
direct measurement of the torus may have been made by N90. As previously
mentioned, north-south scans by N90 measure the 11.2 $\micron$ emitting
region to be 0\farcs16$\pm 0$\farcs04 or $\thicksim 10$ pc. Based on our
``unresolved'' (PSF) component we therefore place an upper limit on the
mid-IR contribution of a dusty torus of $\lesssim 73\%$ of the total
emission at 10.8 $\micron$ and 18.2 $\micron$. This represents the maximum
contribution from a dusty torus and does not rule out contribution to the
``unresolved ''mid-IR emission from a self-absorbed synchrotron source.

Observations of neutral HI and molecular H$_{2}$ by Mundell et al. (1995)
and Fernandez et al. (1999) respectively, show evidence of a gaseous disk up
to 2\farcs5 (160 pc) across which may be associated with the torus. This
disk is located in approximately a north-south direction and may consist of
an ``onion-skin'' morphology as discussed by Pedlar et al. (1998). In this
model the gaseous torus contains several layers (see Figure 5). The
innermost ring consists of ionized gas followed by a ring of neutral HI gas
surrounded by a ring of molecular H$_{2}$. In Pedlar's ``onion-skin'' model
the authors further expand on the subject of anisotropy in NGC 4151 as
discussed by Penston et al. (1990). Using observations of free-free
absorption detected at 73 cm and 18 cm in conjunction with observations of
HI by Mundell et al. (1995) they estimate the ionizing flux incident on the
torus. Assuming a simple Str\"{o}mgren model they calculate that the
ionizing flux in the plane of the torus is between $\thicksim $10 - 40 times
less than seen from Earth or $\thicksim $ 100 - 500 times less than seen in
the NLR as modelled by Penston et al. (1990).

To test the validity of this model we can use the dust equilibrium equation
from Section 4.3. Given values for temperature and dust grain size discussed
above we can calculate the size of the mid-IR torus as a function of
ionizing luminosity, $L_{UV}$. Color temperature measurements of the core of
NGC\ 4151 show T $\thicksim 185\pm 5$ K. If the dusty torus intercepts the
ionizing luminosity as seen from Earth $\thicksim 10^{10}\ $L$_{\sun}$ and
consists of dust grains similar to those estimated for the NLR ($0.04$ $%
\micron$ graphite or $0.01$ $\micron$ SA silicate), the size of the torus in
the mid-IR would be $\thicksim 0\farcs65$ ($\thicksim 42$ pc). This is
slightly larger than the size of the torus based on our resolution limit of $%
\thicksim 0\farcs53$ - $0\farcs58$ ($\lesssim 35$ pc) and that of the Ruiz
polarimetry model ($\thickapprox $ 30 pc). It is also 4 times larger than
the north-south scans by N90 which measured the 11.2 $\micron$ emitting
region to be 0\farcs16$\pm 0$\farcs04 or $\thicksim 10$ pc. Assuming the N90
mid-IR emission delineates the torus, dust grains in the torus would need to
be $\thicksim $10\ times larger than that found in the NLR. Alternatively,
if the luminosity in the plane of the torus is $\thicksim $10 - 40 $\times $
less than seen from Earth as modelled by Pedlar et al. (1998), the size of
the torus in the mid-IR\ would range between $\thicksim 0\farcs1-0\farcs2$ ($%
\thicksim 7-14$ pc). Although slightly smaller than the Ruiz model, this
size torus would be consistent with our upper limit as well as closely match
the size measured by N90. Thus the ``onion-skin'' model of Pedlar et al.\
(1998) which suggests that the luminosity in NGC\ 4151 may be very weak in
the plane of the torus is roughly consistent with size estimates of the
torus in the mid-IR. However, it should be noted that the results discussed
above derive from simple equations involving Str\"{o}mgren radii and dust
grains at equilibrium temperatures. Due to the increased density of material
associated with the torus as opposed to the NLR,\ a more robust radiative
transfer analysis may be needed to truly understand the anisotropy in NGC\
4151.

\section{Conclusions}

In this paper we have used mid-IR\ imaging and first order radiative
transfer analysis assuming dust grains in thermal equilibrium to study the
central $\thicksim 11\arcsec$ of NGC 4151. Our conclusion are as follows.

\ 

1. We detect extended mid-IR emission at 10.8 $\micron$ and 18.2 $\micron$
in the circumnuclear region of NGC 4151. This emission extends approximately 
$\thicksim 200$ pc ($\thicksim $ $3.5\arcsec$) at a P.A. $\thicksim 60\degr$
correlating with the NLR\ region as seen in [OIII] $\lambda $5007 by Evans
et al. (1993) and Kaiser et al. (2000) using HST.

2. With the PSF\ scaled to 100\% of the peak of NGC 4151 we measure limits
to the unresolved and resolved components of $\leq 73\%$ and $\geq 27\%$
respectively.

3. Mid-IR line emission contributes $<10\%$ of the extended emission at 10.8 
$\micron$ and 18.2 $\micron$. The lack of any detectable PAH\ emission also
shows that star formation is weak in the circumnuclear region.

4. Assuming that the luminosity in NGC 4151 is anisotropic ($\thicksim
13\times $), the extended mid-IR emission in NGC 4151 is consistent with
thermal re-radiation from dust grains in the NLR heated by a central engine.

5. We place an upper limit on the size of the torus in the mid-IR of $%
\lesssim 35$pc consistent with the measurements of N90, and Ruiz et. al.
(2002). This results in an upper limit to the mid-IR contribution from a
dusty torus in NGC 4151 of $\leq 73\%$ of the total emission at 10.8 $%
\micron
$ and 18.2 $\micron$ based on our unresolved (PSF) component.

6. Mid-IR measurements of the proposed torus by N90 as well as upper limits
derived from this paper are roughly consistent with the ``onion-skin'' model
of Pedlar et al.\ (1998). In this model, ionizing photons in the plane of
the torus may be $\thicksim $10 - 40 times less than seen from Earth.

\acknowledgments
We would like to thank the Florida Space Grant Consortium for funding which
led to the completion of this work as well as engineer Chris Carter who
provided invaluable support while these observations were taken at Gemini
North.

\clearpage

\begin{deluxetable}{lccc}
\tabletypesize{\footnotesize}
\tablecaption{NGC 4151 Flux Density Measurements. \label{tbl-1}}
\tablewidth{0pt}
\tablehead{
\colhead{Description} &\colhead{Filter} &\colhead{Aperture} & \colhead{Flux Density}} 
\startdata
Total  & N-Band & 4.5$\arcsec$ &  1874$\pm 52$ \tablenotemark{a} \\
Unresolved ($\leq 73\%$) & " & \tablenotemark{b} &  $\leq$1368$\pm 38$  \\
Extended    ($\geq 27\%$) & " & \tablenotemark{b} &  $\geq$506$\pm 14$  \\
\cutinhead{Long Wavelength}
Total  & IHW18& 4.5$\arcsec$ &  4386$\pm 241$\tablenotemark{a}  \\
Unresolved ($\leq 73\%$) & " & \tablenotemark{b} &  $\leq$3202$\pm 176$  \\
Extended    ($\geq 27\%$) & " & \tablenotemark{b} &  $\geq$1184$\pm 65$  \\
\enddata

\tablenotetext{a}{All flux densities are color corrected and in units of mJy. Errors in flux density are dominated by uncertainty in calibration
 ($\pm$ 2.5\% at N-band and $\pm$ 5\% at IHW18) but also include a small statistical error based on the aperture size.}
\tablenotetext{b} {Flux density measurements were performed by scaling the PSF star
$\beta$ Gem to 100\% of the peak of NGC 4151 and subtracting off to find the contribution from the resolved and unresolved component}

\end{deluxetable}

\clearpage

\begin{figure}[tbp]
\plotone{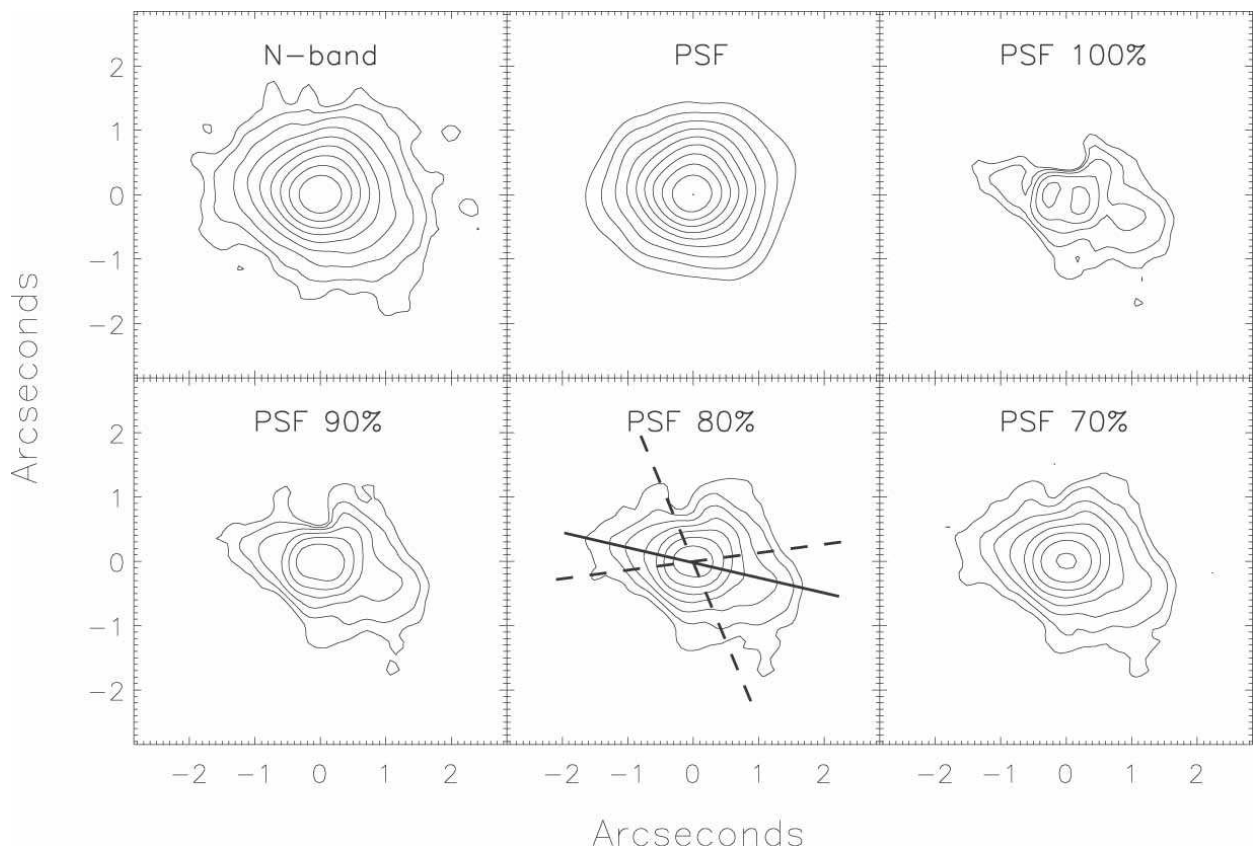}
\caption{N-band images of the central $\sim $ 6$\arcsec$ of NGC 4151. All
images are smoothed with a $\sim $ 0.25$\arcsec$ gaussian filter to enhance
low level emission and scaled logarithmically. The lowest contour represent
the 3 $\protect\sigma $ level of the smoothed data (0.086 mJy). The next
image shows the PSF star $\protect\beta $ Gem scaled to the same level as
NGC 4151 for comparison. The next four images show the residuals of NGC 4151
after subtraction of the PSF (unresolved component) scaled to 100\%, 90\%,
80\%, and 70\% of the peak height. In the 80\% image dashed lines delineate
the edges of the ionization region as observed by Evans et. al. (1993) while
the solid line represents the radio jet axis. With the peak scaled to the
same height as NGC 4151 (100\%), the unresolved component represents $\sim $
73\% of the total emission detected at 10 $\micron$.}
\label{fig1}
\end{figure}

\clearpage 

\begin{figure}[tbp]
\plotone{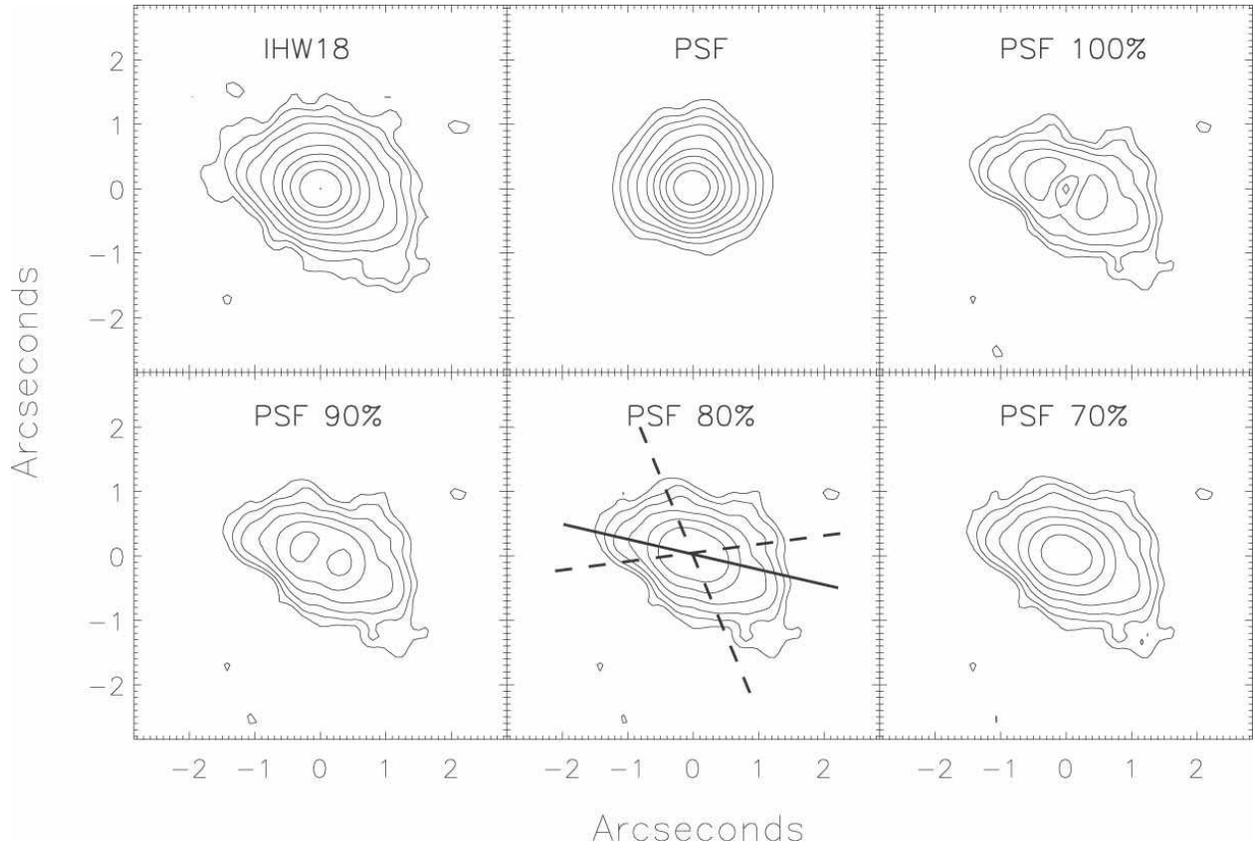}
\caption{IHW18 (18 $\micron$) images of the central $\sim $ 6$\arcsec$ of
NGC 4151 in the same format as Figure 1. The lowest contour represents the 3 
$\protect\sigma $ level of the smoothed data (0.46 mJy). With the peak
scaled to the same height as NGC 4151 (100\%), the unresolved component
represents $\sim $ 73\% of the total emission detected at 18 $\micron$.}
\label{fig2}
\end{figure}

\clearpage 

\begin{figure}[tbp]
\plotone{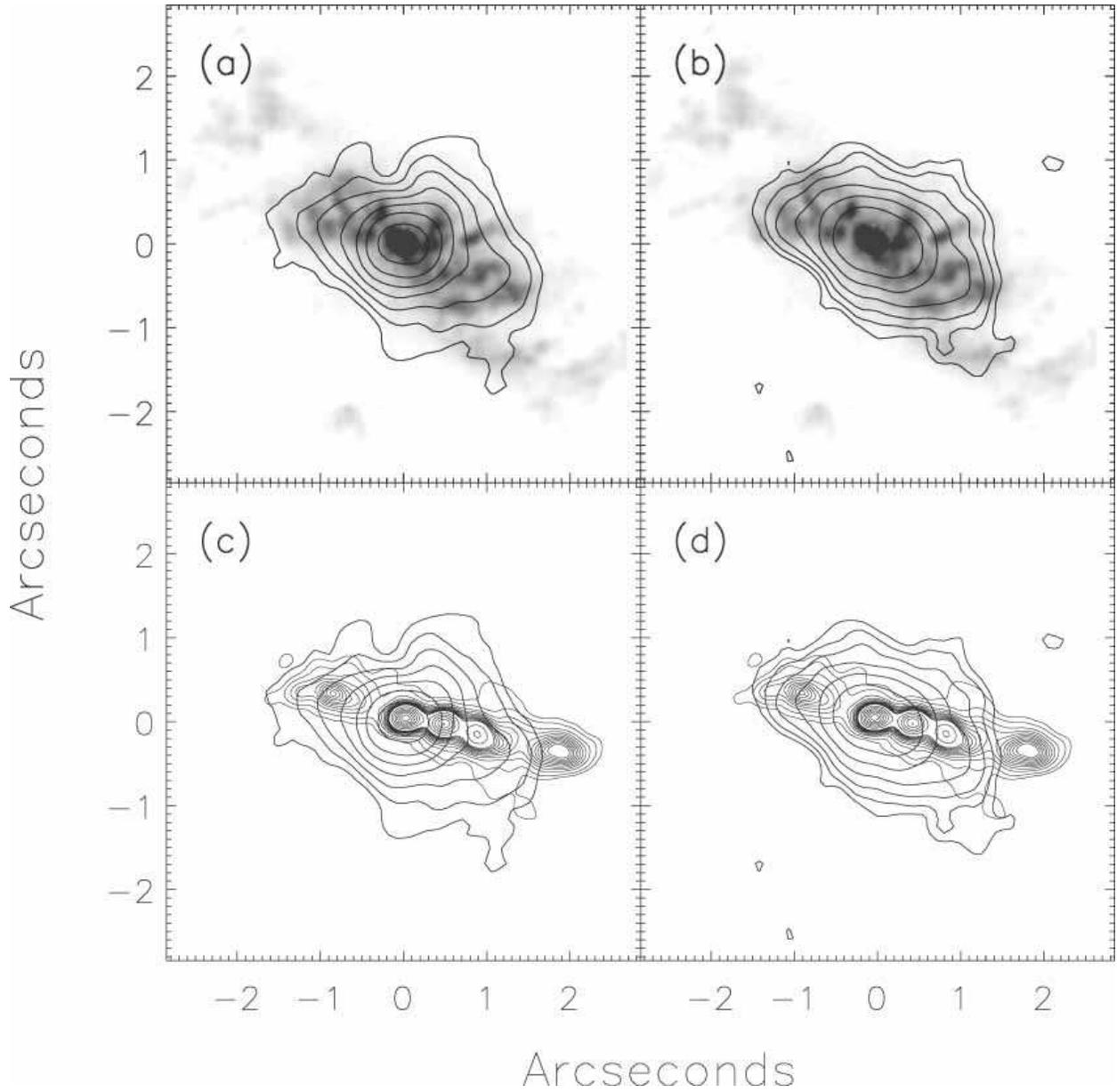}
\caption{Contours represent the extended emission at N and IHW18 after PSF
subtraction (with PSF scaled to 80\% of the peak). Images (a) and (b) show
the extended N and IHW18 emission respectively overlaid on the HST [OIII]
ionization region as observed by Kaiser et al. (2000). Image (c) and (d)
show the same N and IHW18 emission overlaid on the radio jet as observed at
18 cm by Pedlar et al. (1998). In all images the peak emission in the radio and
[OIII] are aligned to correspond with the peak in the mid-IR.}
\label{fig3}
\end{figure}
          
\clearpage 

\begin{figure}[tbp]
\plotone{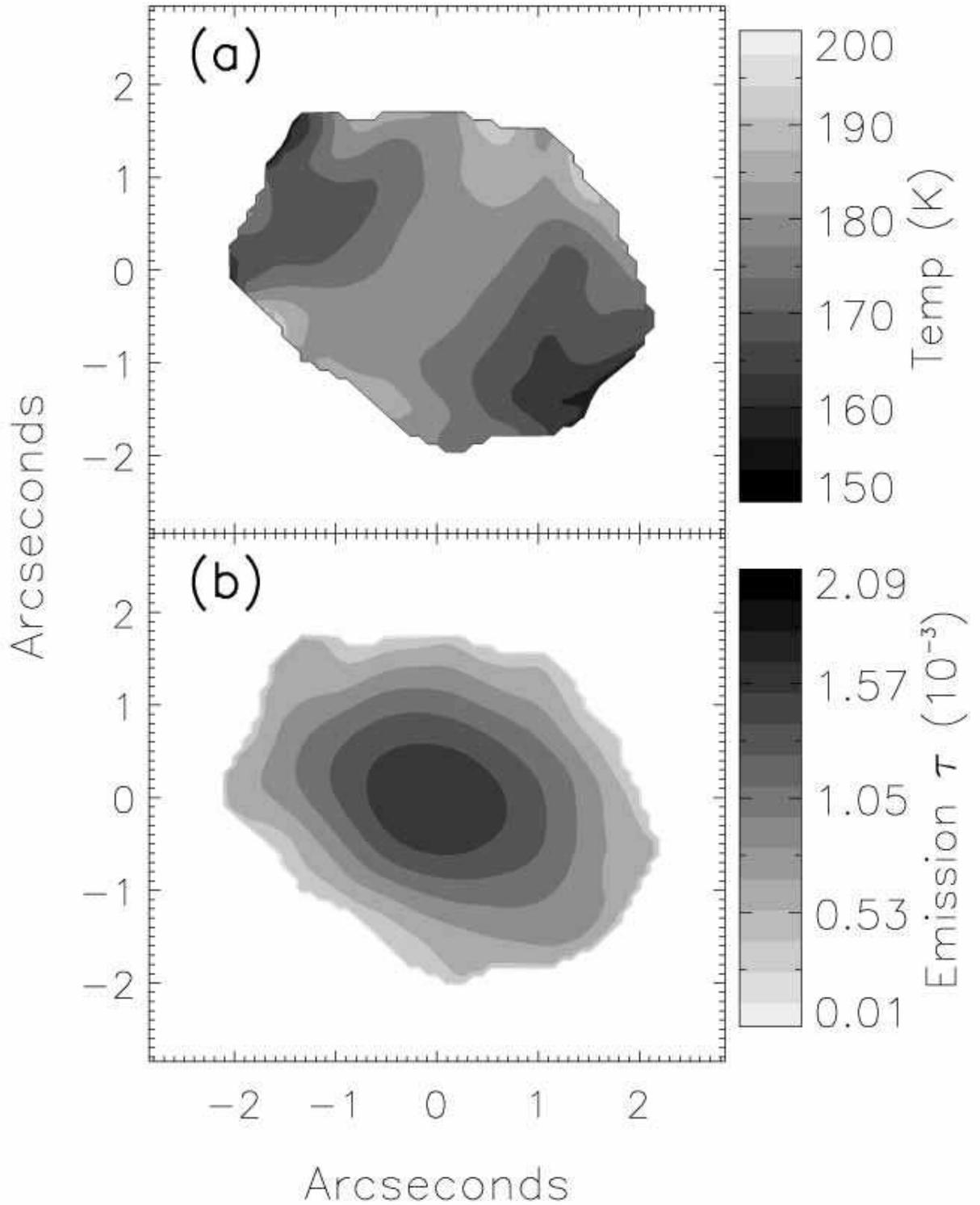}
\caption{Temperature (a) and emission optical depth map (b) of the central $%
\sim $ 6$\arcsec$ of NGC 4151. Temperature peaks along the very outer edges
are erroneous and most likely due to low signal-to-noise. }
\label{fig4}
\end{figure}

\begin{figure}[tbp]
\plotone{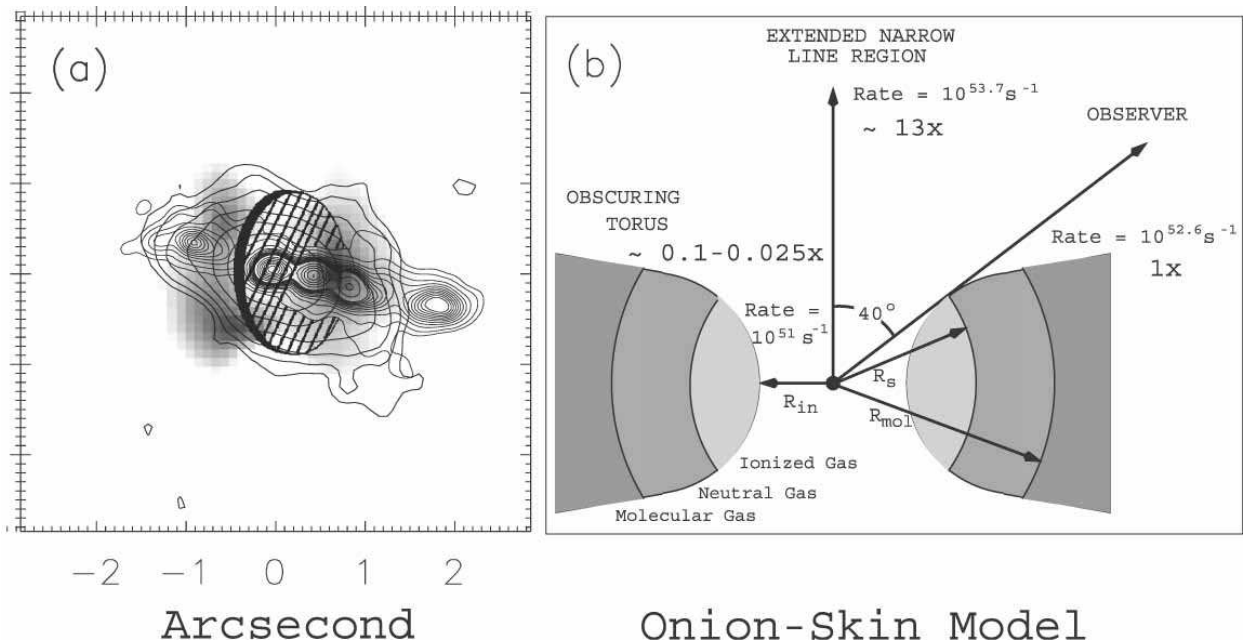}
\caption{Image (a) shows four sets of data. First is the proposed HI disk from 
Mundell et al. (1995) (cross-hatched disk). Second, is the molecular H$_{2}$ 1-0 S(1) ring as 
observed by Fernandez et al. (1999) (grey-scale). Both these images are overlaid 
on the 18 cm radio image from Pedlar et al. (1998) (narrow contours). Finally, 
the larger contours represent our 18.2 $\micron$ image from Figure 2 with the 
subtracted PSF scaled to 80\%. Image (b) shows the ``onion-skin'' model of the gaseous 
torus from Pedlar et al. (1998)(their Figure 6). The rate values are the ionizing photons per second as 
calculated from Penston et al. (1990) (NLR and Observer) and Pedlar et al. (1998) (torus). 
The rate of 10$^{51}$s$^{-1}$ is based on the lower limit from Pedlar et al. (1998). The
upper limit is  $\sim $ 3 times greater, resulting in a range of ionizing flux 0.1-0.025 times
as great as observed from Earth. }
\label{fig5}
\end{figure}

\end{document}